\documentclass[aps,prl,reprint,superscriptaddress]{revtex4-1}

\usepackage{amsmath}
\usepackage{amssymb}
\usepackage{graphicx}

\begin{document}

\title{Coupled charge and spin dynamics in high-density ensembles of nitrogen-vacancy centers in diamond}

\author{R. Giri}
 \email{E-mail: rakshyakar.giri@iit.it}
 \affiliation{Center for Neuroscience and Cognitive Systems, Istituto Italiano di Tecnologia, Corso Bettini 31, Rovereto, 38068, Italy}%
 
\author{F. Gorrini}%
\affiliation{Center for Neuroscience and Cognitive Systems, Istituto Italiano di Tecnologia, Corso Bettini 31, Rovereto, 38068, Italy}%
 \affiliation{Dipartimento di Fisica, Universit\`a degli Studi di Trento, via Sommarive 14, 38123, Povo, Italy}
 
 \author{C. Dorigoni}
\affiliation{Center for Neuroscience and Cognitive Systems, Istituto Italiano di Tecnologia, Corso Bettini 31, Rovereto, 38068, Italy}%

\author{C. E. Avalos}
\affiliation{Institut des sciences et ing\'enierie chimiques, Ecole Polytechnique F\'ed\'erale de Lausanne, 1015, Lausanne, Switzerland}%

\author{M. Cazzanelli}
\author{S. Tambalo}
\author{A. Bifone}
\affiliation{Center for Neuroscience and Cognitive Systems, Istituto Italiano di Tecnologia, Corso Bettini 31, Rovereto, 38068, Italy}%

\date{\today}

\begin{abstract}
We studied the spin depolarization of ensembles of nitrogen-vacancy (NV) centers in nitrogen-rich single crystal diamonds. We found a strong dependence of the evolution of the polarized state in the dark on the concentration of NV centers. At low excitation power, we observed a simple exponential decay profile in the low-density regime and a paradoxical inverted exponential profile in the high-density regime. At higher excitation power, we observed complex behavior, with an initial sharp rise in luminescence signal after the preparation pulse followed by a slower exponential decay. Magnetic field and excitation laser power-dependent measurements suggest that the rapid initial increase of the luminescence signal is related to recharging of the nitrogen-vacancy centers (from neutral to negatively charged) in the dark. The slow relaxing component corresponds to the longitudinal spin relaxation of the NV ensemble. The shape of the decay profile reflects the interplay between two mechanisms: the NV charge state conversion in the dark and the longitudinal spin relaxation. These mechanisms, in turn, are influenced by ionization, recharging and polarization dynamics during excitation. Interestingly, we found that charge dynamics are dominant in NV-dense samples even at very feeble excitation power. These observations may be important for the use of ensembles of NV centers in precession magnetometry and sensing applications.
\end{abstract}

\maketitle

Nitrogen-vacancy (NV) centers in diamond have been explored in recent years for  sensing applications (magnetic field, electric field, and  temperature) \cite{Schirhagl2014, Dolde2011, Rondin2014, Chipaux2015, Tzeng2015, Rudnicki2016, Tetienne2016}, quantum information processing \cite{Childress2013, Alleaume2004}, and as a source of hyperpolarization of ${^{13}}$C, ${^{14}}$N nuclei \cite{Fischer2013PRL, Scheuer2016, Fischer2013PRB}, nitrogen donor spins \cite{Loretz2017} in diamond and also to the nuclei outside the diamond \cite{Abrams2014}. 

The detection sensitivity of these systems scales  as $\sqrt{n}$, where $n$ is the number of negatively charged NV centers (NV$^{-}$) \cite{Wolf2015}. Therefore, for sensing applications where sensitivity is a critical factor, ensembles of NV$^{-}$ centers are preferred over single NV$^{-}$ centers.  In order to achieve high densities of NV centers, vacancies are generated in nitrogen-rich diamond using ion or electron irradiation followed by annealing at high temperatures (\textgreater 800 $^{\circ}$C). In addition to NV defects, irradiation and post-irradiation annealing can create deep level trap states such as neutral NV centers (NV$^{0}$), or  charged nitrogens (N$^{+}$)   as well as divacancies \cite{Deak2014}. These deeps level trap states influence the photophysics and spin relaxation of NV centers \cite{XDChen2015, JayakumarH2016, ChenJ2017}. 

The spin relaxation behavior of ensembles of NV centers in diamond has been studied as a function of NV density, magnetic field and temperature \cite{Jarmola2012, Mrozek2015}. At low NV density, the spin relaxation rate was found to be strongly dependent on temperature, suggesting phonon-mediated relaxation as the dominant mechanism \cite{Jarmola2012}.  At high density, dipole-dipole interactions were shown to drive the relaxation of NV center polarization \cite{Choi2017}. Moreover, additional spin relaxation mechanisms have been proposed in dense ensembles of NV centers, including the formation of fast relaxing centers which could depolarize the whole NV center ensembles \cite{XDChen2015, Choi2017}.  It was hypothesized that the fast relaxing centers are formed as a result of charge dynamics:  ionization of NV centers (NV$^{-} \rightarrow$ NV$^{0} $) through a two-photon process (under intense excitation) \cite{Waldherr2011, Aslam2013}, and subsequent recharging (NV$^{0} \rightarrow$ NV$^{-} $) through electron hopping between defect sites \cite{Mott1956}. Therefore, initialization with a weak laser pulse is preferred to avoid ionization-induced depolarization. Nevertheless, a weak laser pulse still can ionize the NV centers indirectly: ionization of neutral nitrogens (N$^{0}$) through a single photon process and subsequent tunneling of electrons from photoexcited NV$^{-} $ to a nearby N$^{+} $. The efficiency of this process largely depends on the presence of nitrogens \cite{Manson2005}. However, charge state conversion of dense ensembles of NV centers at a low light level, and its influence on spin dynamics is not well understood. In all of the reported experiments, the spin relaxation in the dark consisted of one component that could be fit with a simple or stretched exponential function.  

Here, we have investigated the longitudinal spin relaxation ($T_{1}$) of ensembles of NV centers in a single crystal yellow diamond ($\approx$ 200 ppm of substitutional nitrogen and $\approx$  10 ppb of NV centers) and in a purple diamond with the same concentration of nitrogen, but much higher (\textgreater 1000 fold) concentration of NV centers. At a low excitation power level, we observed an exponential decay of the luminescence signal in the yellow diamond, consistent with previous literature. Surprisingly, an inverted exponential profile was found in the purple diamond, with a sharp growth in luminescence signal following the initialization pulse even at very feeble excitation power. We have investigated the origin of this behavior by measuring longitudinal spin relaxation at various magnetic fields and for different excitation laser powers. We propose a model including the effects of charge and spin dynamics to account for the unusual behavior observed at very high densities of NV centers. We find that the shape and characteristic times of the relaxation curve depend critically on the interplay between these two mechanisms.

\section*{Results}
\begin{figure}
\centering
  \includegraphics[width=1\columnwidth]{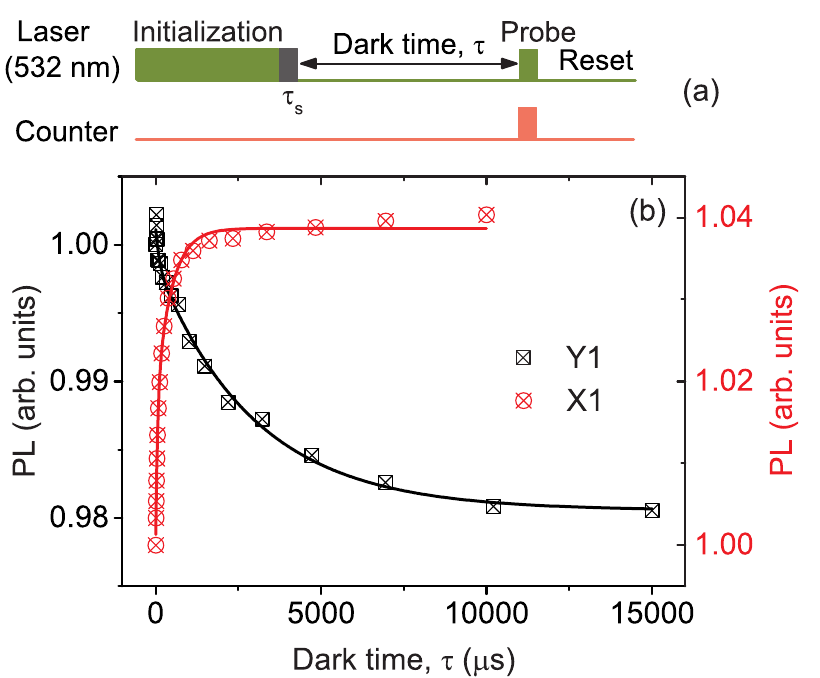}
  \caption{(a) Pulse sequence used to study spin relaxation in the dark. Laser pulses of 500 $\mu$s duration were used to initialize the NV centers, and probe pulses of 1 $\mu$s duration were used to read out the spin-state after variable dark time. The dark time $\tau$ excludes the additional dark time $\tau_s$=1 $\mu$s introduced right after the initialization pulse to allow the spin populations in the singlet states to relax into the $m_s=0$ state.  A reset time of 100 $\mu$s was introduced to enable the system to reach equilibrium. (b) Spin depolarization in the dark: Simple exponential decay profile in sample Y1 and an inverted exponential growth profile in  sample X1. Solid lines are fits with Eq. (1).}
  \label{Fig_1}
\end{figure}

The NV defect ground state is a spin triplet with a zero field splitting of 2.87 GHz between the ${m_s}=0$ and ${m_s}=\pm1$ sublevels. Optical pumping leads to an efficient spin polarization into the ${m_s}=0$ sublevel due to spin dependent intersystem crossing towards an intermediate singlet state. The lifetime of the metastable singlet state varies from 460 $ns$ at 4 K to  150 $ns$ at 450 K \cite{Manson2006, Acosta2010, Robledo2011}. The spin state can be read out through spin-dependent photoluminescence. After initialization into the ${m_s}=0$ state, the NV defect relaxes in the dark to a thermal equilibrium state which is a mixture of  ${m_s}=0, \pm1$ states.  The pulse sequence used for the measurement of the spin-relaxation time is shown in Fig. \ref{Fig_1}(a). The NV centers are excited by a green (532 nm) laser for 500 $\mu$s and probed for 1 $\mu$s after variable delay time $\tau$. An additional dark time $\tau_s$=1 $\mu$s was introduced right after the initialization pulse to allow populations in the metastable singlet state to decay towards the ${m_s}=0$ state.  We introduced a reset time of 100 $\mu$s after the probe pulse to allow the NV centers to reach charge equilibrium. The preparation pulse initializes the NV centers into the $m_{s}=0$ spin state, and relaxation of these NV centers in the dark was studied as a function of $\tau$. All of the experiments reported here were performed at room temperature, and no microwaves were applied. The earth's field was not compensated.

Fig. \ref{Fig_1}(b) shows the spin depolarization in both the samples used in this study, obtained with 200 $\mu$W of excitation laser power at B=0, where B is the external magnetic field.  We observe that the depolarization curves are qualitatively different, with an exponential decay in the low NV density sample  (decay time of the order of $ms$), and a sharp increase in luminescence with a time-scale on the order of 100 $\mu$s in the high NV density sample. These decay profiles and time scales indicate that two distinct mechanisms may determine signal evolution in the two samples. The simple exponential decay profile and time scale on the order of $ms$ are similar to the reported longitudinal spin relaxation of NV ensembles of similar concentration \cite{Jarmola2012, Mrozek2015}. However the sharp increase in the luminescence signal is quite unusual and has not been reported in the literature.

\begin{figure}
 \centering
  \includegraphics[width=1\columnwidth]{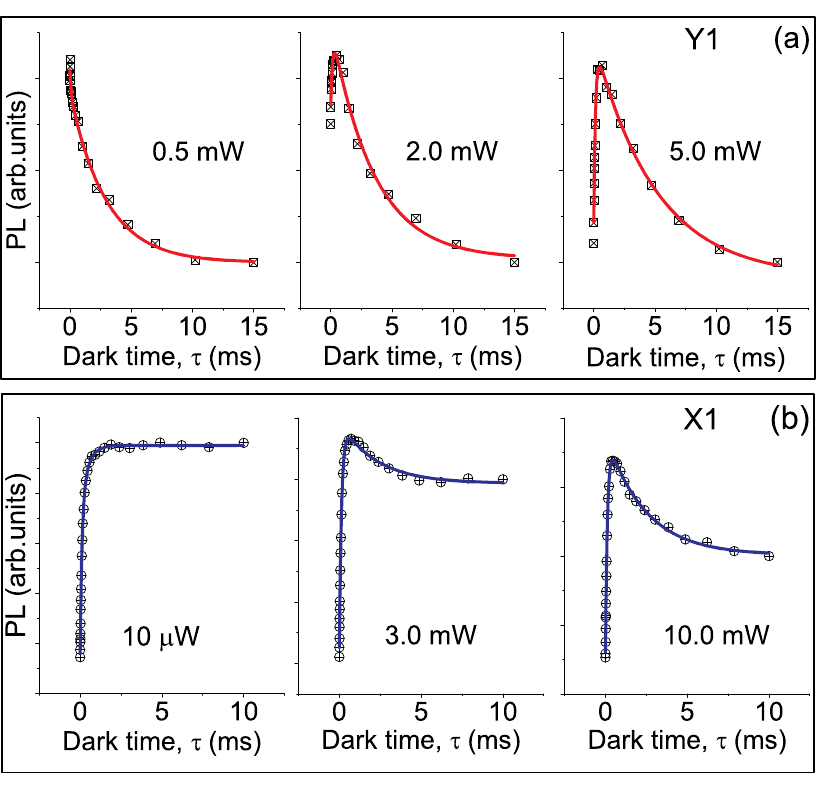}
  \caption{Effect of excitation laser power. Spin relaxation in the dark at various excitation laser power for sample Y1 (a), and sample X1 (b). In the yellow diamond (Y1), at low power, the decay profile is a single exponential, and a fast rising component develops and grows in amplitude as a function laser power. In the purple diamond (X1), at low power, the luminescence signal increases sharply following initialization, and a slow relaxing component builds up with increasing laser power. At high power, the decay profile appears to be qualitatively similar in both samples.}
  \label{Fig_2}
\end{figure}

The spin depolarization curves measured  as a function of excitation laser power are shown in Fig. \ref{Fig_2}. In the low NV density regime (Sample Y1), the decay profile deviates from the initially observed simple exponential type relaxation, and a fast-rising component starts to develop and grows in amplitude with increasing laser power. In the high NV density regime (sample X1) we observed the reverse: the dynamics associated with the fast-rising component dominates at very feeble laser powers (down to 10 $\mu$W), and a slow decaying component develops with increasing laser power. At high excitation power, the depolarization dynamics is qualitatively similar in both high- and low-density regime: a sharp increase in the luminescence signal following the preparation pulse followed by a slower exponential decay. This evidence further confirms the presence of two distinct mechanisms and suggests that both mechanisms may contribute to the signal evolution in the dark under different conditions. The fast-rising component is unlikely due to relaxation from the metastable singlet states as was reported by Tetienne et al. \cite{Tetienne2013}. The metastable state decay time is $\approx$ 200 $ns$ at 300 K  \cite{Acosta2010}, and in the experiments reported here, 1 $\mu$s additional dark time was introduced right after the initialization pulse to allow the spin populations in the metastable state to relax. Therefore, the sharp rise in the luminescence signal observed here (with a time scale of $\approx$ 100 $\mu$s) must be due to a different mechanism.

The green (532 nm) excitation laser pulse used in this study can ionize the NV centers (NV$^{-} \rightarrow$ NV$^{0} $) through a two-photon absorption process, and this process is more efficient at higher excitation power \cite{Waldherr2011, Aslam2013}. However, it has been suggested that ionization can occur even at low light level via a tunneling process from a photoexcited NV$^{-}$ to a neighbouring substitutional nitrogen N$^{+}$ \cite{Manson2005, Loretz2017}. These ionized NV centers can recharge (NV$^{0}$+N$^{0} \rightarrow$ NV$^{-}$+N$^{+}$) and reach a charge equilibrium state. The recharging process can happen in the dark in the absence of any optical, electrical or thermal excitation, and the mechanism suggested for this is tunneling of electrons among closely spaced NV centers \cite{Choi2017}.  The recharging process could also be due to an impurity conduction process: electrons hopping between NV centers and  nitrogens (N$^{0}$) \cite{Mott1956}. This dynamics largely depends on the concentration and distribution of N$^{0}$. If the concentration of N$^{0}$ is high enough, then availability of electrons to tunnel could facilitate recharging of NV centers in the dark \cite{Manson2005}. A characteristic recharging time of around 100 $\mu$s was found in a sample containing 45 ppm of NV centers \cite{Choi2017}, which is close to the time scale observed in this study. The initial rise in the signal observed in our experiments is possibly related to this recharging process, and from now onwards we will refer to the corresponding time constant as the recharge time, ${T_r}$.

All of the above observations suggest that the decay profile reported in this study could be a result of two competing processes pertaining to charge dynamics (ionization-recharge) and spin dynamics (polarization-relaxation). 

\begin{figure}
\centering
  \includegraphics[width=1\columnwidth] {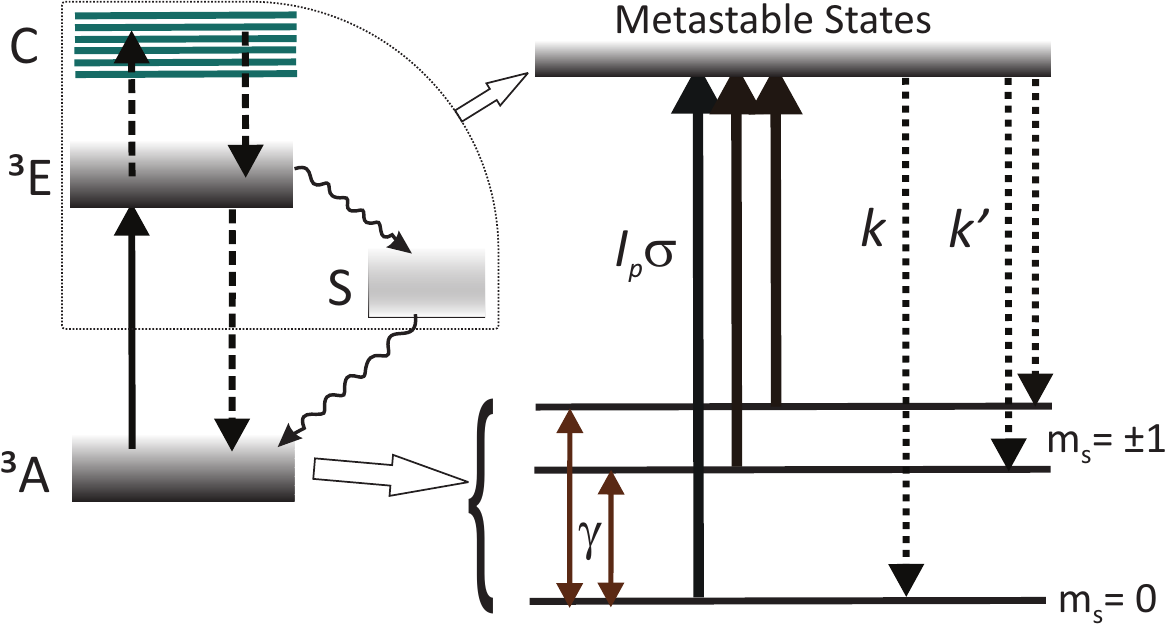}
  \caption{Simplified energy levels of a NV center. ${^{3}}$A and ${^{3}}$E are the ground and excited state triplets respectively. Label C represents the conduction band, and label S includes all the singlet states. In the simplified 4-level scheme, the metastable state includes the excite state, conduction band and the singlet states.}
  \label{Fig_3}
\end{figure}

 We model the polarization-depolarization and ionization-recombination dynamics by evaluating the populations within a simplified  4-level scheme as shown in Fig. \ref{Fig_3}. Ionization-recombination of the NV centers under various light levels also affects the spin populations in the ${m_s}=0$ state. We consider the ground state triplet and a "metastable" level that includes the excited state, the conduction band, and the dark singlet states. 
 
\begin{figure}[h]
\centering
  \includegraphics[width=1\columnwidth] {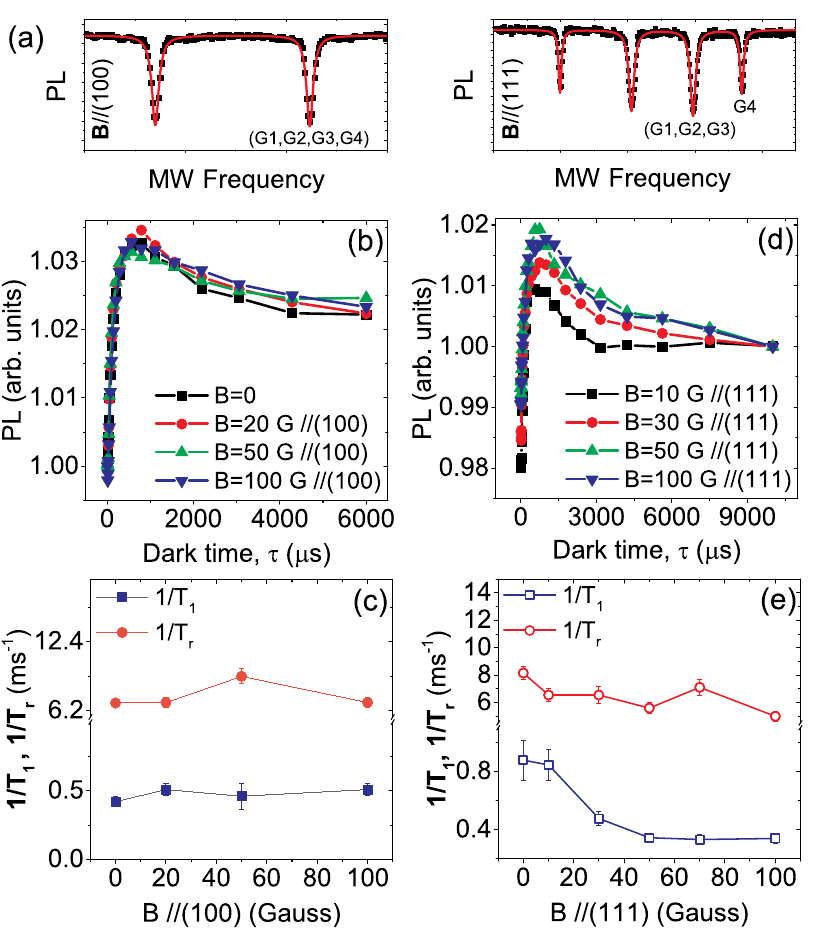}
  \caption{Effect of external magnetic field. (a) Electron spin resonance spectra obtained from sample X1 for the case of magnetic field B=//(100), when all four subgroups of NV centers are resonant (left); and for B aligned along (111) direction (right). In this case one group of NV centers (G4) is separated from the rest three resonant groups (G1, G2, G3). PL decay profile in the dark for different values of B // (100) (b), and  B//(111) (d). The recharging rate (1/$T_{r}$) and  spin relaxation rate  (1/$T_{1}$) plotted as a function of B // (100) (c), and B // (111) (e). The recharging rate is not influenced by external field, whereas the spin relaxation rate does depend on B//(111).}
  \label{Fig_4}
\end{figure}  
 
 We define ${T_1}$ as the relaxation time of the populations in the $\vert {m_s}=0 \rangle$ state, and ${T_r}$  as the recharging time (NV$^{0} \rightarrow$ NV$^{-}$). Solving the rate equations for the populations, we can write the evolution of the photoluminescence (PL) signal as :
 \begin{equation}
I(\tau)=I_{eq} \lbrack1-\alpha(I_p) e^{-\tau/{T_r}}+\beta(I_p)e^{-\tau/{T_1}} \rbrack
\label{Eq1}
\end{equation}
Here ${I_{eq}}$ is the PL at equilibrium, $1/{T_r}=(k+2k^{'}+I_p\sigma)$ is the recharging rate and  $1/{T_1}=(3\gamma+I_p\sigma)$ is the spin relaxation rate; $\gamma$ is the decay rate from $\vert0\rangle$ to $\vert\pm1\rangle$, \textit{$I_p$} is the laser intensity, $\sigma$  is the cross-section for both absorption and ionization, and k, k' are the charge capture rates. During initialization, the recharging rate increases with laser intensity, consistent with previous observation \cite{ChenJ2017}. In the dark (\textit{$I_p$}=0), the recharging rate is solely dependent on the charge capturing rates, and the spin relaxation rate depends on the decay rate $\gamma$. However, the amplitude of charge state conversion ($\alpha$) and the amplitude of spin dynamics ($\beta$) depends on the excitation intensity, rate of charge capture as well as rate of depolarization i.e., $\alpha$($\gamma,k,k^{'},I_p\sigma$) and $\beta$($\gamma,k,k^{'},I_p\sigma$). So, these two processes are not completely independent of each other, and the depolarization dynamics in the dark is determined by the relative weight of these two competing processes. We note that our model does not suggest anything about the physical mechanisms behind the recharging process in the dark.

\begin{figure}
\centering
  \includegraphics[width=1\columnwidth]{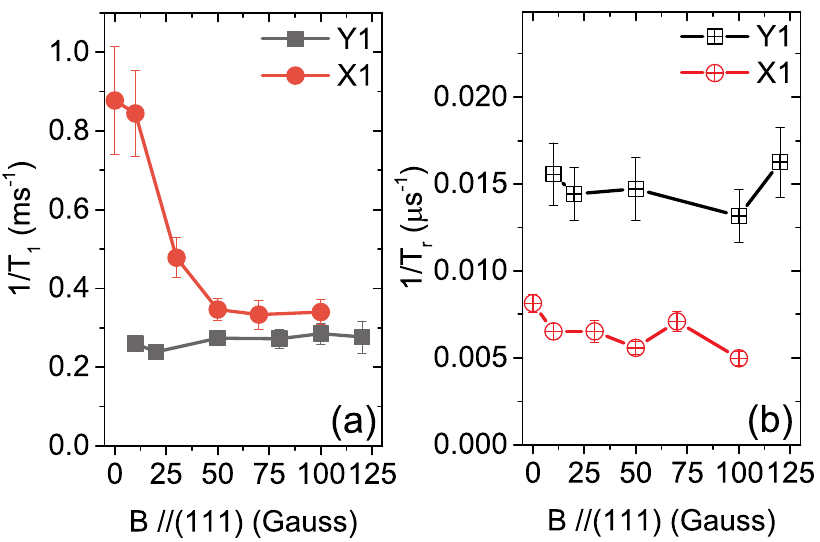}
  \caption{(a) Spin relaxation rate (1/$T_{1}$), and (b) recharging rate (1/$T_{r}$) as a function of B//(111). Spin relaxation rate decreases with increase in B in the high NV density regime (X1), but the recharging rate is independent of B in both the high and low NV density regime.} 
  \label{Fig_5}
\end{figure}

In order to investigate our hypothesis that the shape of the decay profile is a result of two competing processes pertaining to charge dynamics and spin dynamics, we measured spin depolarization as a function of external magnetic field [Fig. \ref{Fig_4}]. Indeed, we expect the two mechanisms to depend differently on magnetic field strength and orientation.  There are four groups of NV centers in diamond corresponding to four different symmetry axis orientations.  At zero magnetic field as well as magnetic field aligned along (100) direction, transition frequencies of all the NV centers overlap and cross-relaxation is maximum. If the magnetic field is aligned along the (111) direction, then one group of NV centers (G4) have transition frequency different from the rest of the three groups (G1, G2, G3) which are still degenerate [Fig. \ref{Fig_4} (a)]. The detuning in transition frequencies can be increased by increasing the magnetic field strength, reducing the cross-relaxation. We observe that the depolarization rate of the slow relaxing component is strongly influenced by the change in magnetic field strength. The depolarization rate is maximum at zero field and decreases slowly as a function of magnetic field [Fig. \ref{Fig_4} (e)]. Therefore, the NV-NV cross-relaxation induced magnetic noise is most likely responsible for the depolarization of the slow relaxing component.  Further validation to this inference comes from the fact that this dynamics is only observed in the high NV density regime [Fig. \ref{Fig_5} (a)].  We can conclusively assign the slow relaxing component to the longitudinal spin relaxation (${T_1}$) of the NV centers. The sharp rising component is not affected by  magnetic interaction, consistent with a charge-driven mechanism [Fig. \ref{Fig_5} (b)].

\section*{Discussion}
\begin{figure}
\centering
  \includegraphics[width=1\columnwidth]{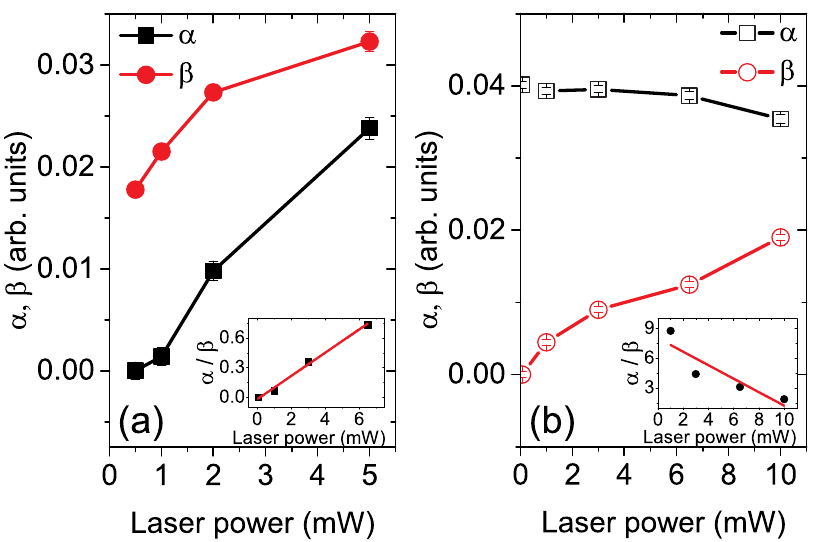}
  \caption{Amplitudes of charge state conversion ($\alpha$), and spin dynamics (polarization-relaxation) ($\beta$)  as a function of excitation laser power in sample Y1 (a), and X1 (b). The solid lines are guide to the eye. Insets: Linear dependence of  ($\alpha/\beta$)  on excitation laser power.}
  \label{Fig_6}
\end{figure}

The dynamics of ionization, recharging and polarization during excitation and the resulting effect on depolarization dynamics in the dark  in the two regimes of low and high density of NV centers can be described as follows:

\textbf{Low NV density regime (Sample Y1):}  In this case, both $\alpha$ and $\beta$ increase linearly with laser power, but the change in $\alpha$ is more prominent [Fig. \ref{Fig_6}(a)]. Indeed, at the lowest power, ionization through two-photon process is inefficient, and the decay profile in the dark is a single exponential reflecting the dominant spin dynamics process during initialization [Fig. \ref{Fig_2}(a)]. As the excitation laser power increases, NV$^{-}$ centers are ionized creating more NV$^{0}$  centers \cite{XDChen2013, BarGill2017}. The laser pulse also ionizes the nitrogen donors \cite{Heremans2009, JayakumarH2016}, and these electrons can be captured by NV$^{0}$  centers to form NV$^{-}$. The rate of ionization and recharging process increases as the laser power increases.  Increasing $\beta$ indicates an increase in the degree of polarization with laser power. When the laser pulse is switched off, the recharging process continues in the dark  until a charge equilibrium is attained. These dynamics are reflected in the bi-exponential decay profile. 

\textbf{High NV density regime (Sample X1):} In this case, $\alpha$ appears to be constant, however $\beta$ increases almost linearly with power [Fig. \ref{Fig_6}(b)]. Charge dynamics dominate even at the lowest power and appears to saturate [Fig. \ref{Fig_2}(b)]. The strong charge dynamics at low laser power is quite unusual. Usually at low power, ionization is minimal because of the inefficiency of two-photon process and spin polarization as high as 90 \% can be achieved \cite{Harrison2006, Howard2006, Felton2009, Waldherr2011}. But here the rate of polarization is  low in comparison to ionization-recharging process. We must note that this sample is highly disordered with more than 200 ppm substitutional N$^{0}$, 10 ppm of NV$^{-}$ centers, and 100 ppm of nitrogens occupying the interstitial sites. The interstitial nitrogens could be charged (N$^{+}$) \cite{Woods1984}, or neutral \cite{Kiflawi1996, Mainwood1999, Felton2009}. It could also contain significant number other deep level trap states such as divacancies \cite{Deak2014}. It may also be possible that NV$^{-}$ centers exist in the form of [NV$^{-}$ $-$ N$^{+}$] pairs rather than as an isolated center \cite{Manson2005}. Even at the lowest light level, the photo-excited NV$^{-}$ centers could lose electrons to the charged nitrogens (N$^{+}$) through tunneling $\left ( \lbrack \textrm{NV}^{-} - \textrm{N}^{+}\rbrack \rightarrow \lbrack \textrm{NV}^{0} - \textrm{N}^{0} \rbrack \right)$\cite{Manson2005, Loretz2017}. Ionization of nitrogen donors (N$^{0}$) is also efficient resulting in abundant  electrons in the conduction band.  These conduction electrons can be trapped by NV$^{0}$ to form NV$^{-}$: $\left( \lbrack \textrm{NV}^{0} - \textrm{N}^{0} \rbrack \rightarrow \lbrack \textrm{NV}^{-} - \textrm{N}^{+} \rbrack \right)$. Therefore, the ionization-recharging process is robust even at low laser power.   As the excitation power increases more N$^{0}$ are ionized and trapping states are filled with electrons resulting in  depletion of the N$^{0}$ reservoir \cite{ChenJ2017}. We assume that the ionization of the electrons trapped in deep level defects is slower than ionization of the donor electrons, so the charge dynamics slows down and reaches an equilibrium resulting in an increase of NV$^{-}$ density. We are in a regime where pumping rate is slower than spin relaxation rate. Thus, increase in laser power results in an increase in the degree of polarization of the NV$^{-}$ centers.  The saturation of charge dynamics during pumping is reflected in the constant amplitude of charge state conversion in the dark ($\alpha$), and increase in the degree of polarization as a function of laser power results in a linear increase in the amplitude of spin dynamics (polarization-relaxation) ($\beta$).

In summary, we studied the spin depolarization of ensembles of NV centers in single crystal diamond. At low excitation power, we observed a simple exponential type decay in the low NV density regime, but a sharp rise in the luminescence signal right after the initialization pulse in the high NV density regime. At higher excitation power, we observed complex behavior, with a sharp rise in luminescence signal followed by a slower exponential decay. Experiments with varying excitation laser power and magnetic field provide evidence that these decay profiles are due to a complex process involving charge and spin dynamics during pumping as well as in the dark. Our analysis indicates that charge dynamics, possibly due to tunneling among a network of closely spaced NV centers, or among NV centers and nearby nitrogen atoms, may be dominant in the high-density regime even at very low laser excitation power. In this context, interstitial nitrogen present in NV dense diamonds may play a significant role, even though a more detailed and systematic investigation of the involvement of these defects is needed.  Our results corroborate previously reported findings that recharging of NV centers can occur in the dark in the absence of any laser or microwave irradiation \cite{Choi2017}. Due to the effects of charge state conversion on spin relaxation measurements, controlling charge dynamics is vital in the application of ensembles of NV centers as sensors or for hyperpolarization of nuclear spins. A deeper understanding of the effect of charge state conversion will have a significant impact on the use of T$_{1}$ based sensing schemes.


\section*{Methods}
\subsection*{Samples}
Two high-pressure, high-temperature (HPHT) single crystal diamonds cut along the (100) crystallographic direction are used in this work and were produced by Element Six. One sample, "yellow" (Y1), is a standard commercial Type Ib HPHT diamond. The other one, "purple" (X1), is a Type Ib HPHT diamond, electron irradiated and annealed to increase the concentration of NV centers. We determined the substitutional nitrogen (N$^{0}$) concentration from the absorption coefficient of the feature at 1135 cm$^{-1}$  in the IR absorption spectrum (not shown here), following the methodology of Woods et al. \cite{Woods1990}. We estimated that both samples contain more than 200 ppm of N$^{0}$.  We observed that the IR absorption spectrum of sample X1 presents a feature at 1450 cm$^{-1}$ which corresponds to interstitial nitrogens \cite{Woods1984, Kiflawi1996}. This feature is usually found in Type Ib diamond irradiated with high fluence and annealed at high temperature. We estimated that sample X1 contains about 100 ppm of interstitial nitrogens. In sample Y1, presence of interstitial nitrogens was not detectable.  From the absorption coefficient of 637 nm line in the UV-Vis absorption spectrum, and using the calibration constants from G. Davies \cite{Davies1999}, we estimated that sample X1 contains $\approx$ 10 ppm of NV$^{-}$.  Comparing the luminescence of the two samples, we determined that the sample Y1 contains $\approx$ 10 ppb of NV$^{-}$.

\subsection*{Experimental setup}
We used a confocal microscope built in-house to study the depolarization of NV center ensembles. A microscope objective with a numerical aperture (NA) of 0.65 was used to focus the excitation laser (532 nm, Coherent Verdi) to a focal spot of around 10 $\mu$m. The fluorescence from the NV centers was collected with the same objective, filtered by a series of low- and high-pass filters and a single photon counting module (Excelitas, SPCM-AQRH-14-FC) detected the luminescence in the range of 620-750 nm. We used a 3-axis Helmholtz coil system for controlling the strength and direction of a static magnetic field.  An acousto-optic modulator (AA Optoelectronics) produced the excitation laser pulses, and a programmable TTL pulse generator (PulseBlaster ESR-PRO) was used to generate the pulse sequences.

\normalsize

\section*{Acknowledgements}
We thank Dr. Michele Orlandi and Prof. Antonio Miotello of IdEA laboratory, University of Trento, for helping us with the measurement of UV-Vis absorption spectra of diamond samples. 

\end{document}